\documentclass{PoS}
\usepackage{epsfig,amsmath,amssymb,graphicx}
\pdfoutput=1

\newcommand{\order}[1]{\mathcal{O}(#1)}
\newcommand{\pr}{\mbox{pr}}
\newcommand{\as}{\mathbf{a}}
\newcommand{\ares}{\mathbf{a}_{res}}
\newcommand{\amarg}{\mathbf{a}_{marg}}

\title{A Bayesian approach to chiral extrapolations}

\ShortTitle{A Bayesian approach to chiral extrapolations}

\author{\speaker{Matthias R.~Schindler}\\
         Department of Physics and Astronomy
         Ohio University\\
         Athens, OH 45701\\
         USA\\
         E-mail: \email{schindle@ohio.edu}}

\author{Daniel R.~Phillips\\
         Department of Physics and Astronomy
         Ohio University\\
         Athens, OH 45701\\
         USA\\
         E-mail: \email{phillips@phy.ohiou.edu}}

\abstract{The determination of low-energy constants from data is an important component of most effective field theory programs, including that of chiral perturbation theory. We propose a novel method based on Bayesian probability theory which allows us to address several shortcomings of the standard approach to parameter extraction. Using a toy-model we argue that the Bayesian approach is ideally suited for the application in effective field theories. We also discuss the application to lattice QCD data.}

\FullConference{6th International Workshop on Chiral Dynamics\\
		 July 6-10 2009\\
		 Bern, Switzerland}

\begin{document}

\section{Introduction}

An effective field theory (EFT) is a low-energy approximation to an underlying theory. It allows for a model-independent description of phenomena at an energy scale $m$ that is much lower than an underlying scale $\Lambda$. The Lagrangian of the EFT is constructed by including all terms that are consistent with the symmetries of the underlying theory. Each of the terms in the Lagrangian is accompanied by a so-called low-energy constant (LEC) that incorporates the effects of high-energy degrees of freedom on the low-energy dynamics. The EFT leads to a perturbative expansion for observables at the low-energy scale if 
the LECs are of order $\order{1}$ in units of the high-energy scale, i.e.~if they are ``natural" with respect to $\Lambda$. In principle, these LECs can be determined from the underlying theory. In practice, however, there are only a few cases in which the LECs can be rigorously derived from the underlying theory, and in all other instances the only model-independent way to determine the LECs is by comparison with experimental data. 

The standard approach to the extraction of LECs from data is to calculate an observable at some given order and then perform a fit of this EFT expression using methods like least squares or maximum likelihood. There are several issues with this approach that we are going to address:

\begin{enumerate}
\item Which order in the EFT expansion should be used to perform the fit?
\item How can the naturalness requirement on the LECs be incorporated?
\item What is the appropriate energy regime to perform the fit? In most cases more data is available for higher energies, but the reliability of the EFT calculation decreases as the energy is increased.
\end{enumerate}
With data sets that include a large number of very precise measurements, these issues are not of any significance. If, however, only limited and imprecise data is available, these issues manifest themselves as sensitivity of the extracted LECs on the way the fit is performed.

In order to avoid the above-mentioned issues we have developed an approach that is based on Bayesian probability theory \cite{Schindler:2008fh}. We argue that  Bayesian methods (for an introduction see e.g.~\cite{Sivia}) are ideally suited for the extraction of LECs. In the Bayesian approach prior knowledge on the parameters can be easily included in the process of estimating these parameters. When combined with the concept of marginalization, applied to the order of the fit function, the derived method resolves the first two issues in the above list. We also show that this method is not sensitive to higher-energy data within certain bounds.

\section{Bayesian probability theory}

Consider a general EFT for which the LECs are denoted by $\as=\{a_i|i=1,\ldots,M\}$. In the following we will restrict the discussion to extracting a subset $\ares$ of these unknown parameters from some given data $D=\{(d_k,\sigma_k)|k=1,\ldots,N\}$, where $d_k$ is an individual measurement at $x_k$ with associated uncertainty $\sigma_k$. We are therefore interested in the probability density
\begin{equation}
\pr(\ares|D),
\end{equation}
where $\pr(A|B)$ denotes the conditional probability density of $A$ given $B$.
Bayes' theorem relates this probability density to the more familiar likelihood $\pr(D|\ares)$,
\begin{equation}
\pr(\ares|D)=\frac{\pr(D|\ares)\pr(\ares)}{\pr(D)}.
\end{equation}
Here, $\pr(\ares)$ is the so-called prior which incorporates any information available on the parameters prior to analysis of the data. The denominator can be obtained from the requirement that $\pr(\ares|D)$ be normalized. The prior information we wish to include is the assumption of naturalness of the parameters. However, the notion of ``naturalness'' is not  strictly defined. Here we employ the principle of maximum entropy to motivate a prior of the form~\cite{Schindler:2008fh}:
\begin{equation}\label{priorpdf}
\pr(\as|M,R)=\left(\frac{1}{\sqrt{2\pi} R}\right)^{M+1} \exp\left(-\frac{\as^2}{2R^2}\right).
\end{equation}
Note that we have introduced several additional parameters: $\as=(\ares,\amarg)$ denotes the complete set of LECs at a given order, including the higher-order LECs $\amarg$ that we do \emph{not} wish to extract, $M$ is related to the order of the EFT calculation,\footnote{In general, $M$, the number of LECs, is not identical to the order of the calculation.} and $R$ is a parameter that encodes the definition of naturalness as chosen here. Thus, while we have succeeded in defining the prior, this has come at the price of the introduction of these additional parameters. Since we are not interested in the exact values of these parameters and, in fact, one of our aims was to avoid having to fix the value of $M$, we apply marginalization to eliminate these ``nuisance'' parameters. The general marginalization description is given by
\begin{equation}
\pr(A|C)=\int dB \,\pr(A,B|C),
\end{equation}
that means unwanted parameters are integrated out. We apply marginalization to the higher-order LECs $\amarg$, the order of the EFT calculation and the ``naturalness parameter'' $R$. This last marginalization thus takes  into account the uncertainty in the definition of naturalness. The final probability density is given by (for a derivation see Ref.~\cite{Schindler:2008fh})
\begin{equation}\label{finalpdf}
\pr(\ares|D)=\sum_M \int dR\, \int d\amarg\, \frac{\pr(D|\as,M)\pr(\as|M,R)\pr(M)\pr(R)}{\pr(D)}\, .
\end{equation}
Since Bayes' theorem was employed several times in the derivation of Eq.~(\ref{finalpdf}) we are forced to introduce priors for $M$ and $R$. We do not assume any particular knowledge of these parameters. However, since $M$ is a ``location parameter'' and $R$ is a ``scale parameter'' we use different priors. The prior for $M$ is a constant, while $\pr(R)=\frac{1}{R}$ (see Ref.~\cite{Schindler:2008fh} for more details). The parameters and the associated uncertainties are determined from the first and second moments of the pdf,
\begin{align}
\langle a_i \rangle &= \int d\ares\, a_i \pr(\ares|D), \\
\sigma_{a_i}^2 &=\langle a_i^2 \rangle -\langle a_i \rangle^2.
\end{align}

\section{Application to a toy problem}

In order to demonstrate the advantages of our proposed method we consider an application to a toy problem. We generate pseudo-data using the function
\begin{equation}\label{Ex:Func}
g(x)=\left(\frac{1}{2}+\tan\left(\frac{\pi}{2} x\right) \right)^2
\end{equation}
for $x \ge 0$. Our aim is to extract the first two coefficients $a_0,a_1$ of a polynomial
\begin{equation}
f(x)=\sum_{j=0}^P a_j x^j
\end{equation}
from the pseudo-data, where $P$ denotes the order of the polynomial. The function $g(x)$ might not have any direct physical application, but it exhibits a number of features that are common in EFT applications. $g(x)$ is non-analytic in $x \in \mathbb{R}$, but for $x<1$ it can be approximated to arbitrary precision by a power series. The first few terms in this power series are given by
\begin{equation}
g(x)\approx 0.25+ 1.57 x + 2.47 x^2 + 1.29 x^3 + 4.06 x^4 + \cdots.
\end{equation}
The coefficients of at least the first ten terms are ``natural'', however, their magnitude is not decreasing for increasing order.

\begin{figure}
\begin{center}
\includegraphics{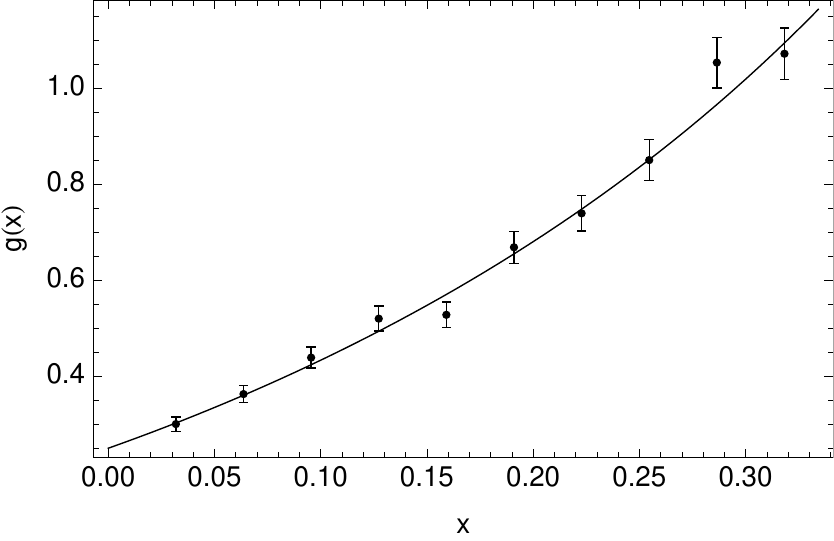}
\end{center}
\caption{\label{Fig:data}Generated artificial data. The solid line is the function $g(x)$.\label{Ex:dataMax05Err5}}
\end{figure}

The pseudo-data we wish to analyze, covering the range $0<x\le 1/\pi$ are shown in Fig.~\ref{Fig:data}. The prior information available is that the data are normally distributed (which reduces the problem to a minimum $\chi^2$ one in the standard approach) and that the coefficients of the polynomial are $\order{1}$.  The results of a standard least-squares fit at various orders of the polynomial, which does not take into account the information on the naturalness of the parameters, are shown in Tab.~\ref{Tab:Standard}. While the quadratic fit reproduces the underlying values of $a_0$ and $a_1$ reasonably well and with a relatively low $\chi^2$, without knowledge of the underlying values it might be difficult to decide why the quadratic fit should be preferred. One should also note the lack of convergence, especially for $a_1$, as one goes to higher orders and the fast growth of the uncertainties. An experienced practitioner might be able to discern which of the various fits to trust most, however our aim is to eliminate the need for this post-analysis judgement.
\begin{table}
  \centering
$$\begin{array}{c|c|c|c}
P & \chi^2/d.o.f. & a_0 &  a_1  \\ \hline
1 & 2.23 & 0.203  \pm  0.014 & 2.51  \pm  0.10  \\
2 & 1.06 & 0.260  \pm  0.022 & 1.31  \pm  0.39  \\ 
3 & 1.13 & 0.235  \pm  0.038 & 2.14  \pm  1.08  \\ 
4 & 1.13 & 0.177  \pm  0.067 & 4.76  \pm  2.70  \\ 
5 & 0.99 & 0.327  \pm  0.133 & -3.56  \pm  6.94 \\ 
6 & 1.32 & 0.314  \pm  0.297 & -2.73  \pm  18.5 \\ 
7 & 1.47 & 1.05   \pm  0.792 & -56.3  \pm  56.5 \\ 
\end{array}$$
  \caption{\label{Tab:Standard}Fit result for standard $\chi^2$ approach.}  
\end{table}

We now apply our method based on the use of Bayes' theorem and marginalization to the data. The naturalness of the parameters is included in the analysis with the use of the prior of Eq.~(\ref{priorpdf}). We marginalize over the polynomial order from $P=2$ to $P=8$. For the ``naturalness parameter'' $R$ we choose $R=0.1-10$. We find
\begin{align}
  \label{eq:4}
  a_0&=0.246\pm0.021, \\
  a_1&=1.63\pm0.37,
\end{align}
in good agreement with the underlying values. We have avoided the need to choose a specific order for the fit; instead the uncertainties in the results for $a_0$ and $a_1$ include contributions from the marginalization over $P$. And while the results are influenced by our inclusion of the ``naturalness prior'', the lack of exact knowledge of $R$ again contributes to the final uncertainties of the parameters via marginalization. We therefore believe that our method not only leads to improved extraction of the parameters of interest, but also includes some of the uncertainties related to such an extraction in a more systematic way than the standard approach.

We have performed an analogous analysis with a different data set that contains the same number of data points, but for which $0<x\le 2/\pi$. The result for the standard $\chi^2$ fit are shown in Tab.~\ref{Tab:Standard2}. With more data points closer to the radius of convergence the problems of the standard approach are exacerbated. While the fourth-order fit gives results not too far from the underlying values, without knowledge of these ``true'' values it is not clear which result to trust.
In our Bayesian approach, again choosing $P=2-8$ and $R=0.1-10$, we find
\begin{align}
  a_0&=0.241\pm0.048,\\
  a_1&=2.23\pm0.74.
\end{align}
These values are again in agreement with the underlying values, and reproduce them much better than the standard $\chi^2$ results. We consider it a strength of our method that the results are not as sensitive to high-x data, allowing for the use of larger data sets.

\begin{table}
\begin{center}
\begin{tabular}{c|c|c|c}
$M$ & $\chi^2/d.o.f.$ & $a_0$ & $a_1$ \\ \hline
2 & 5.35 & 0.392 $\pm$ 0.033 & -0.387 $\pm$ 0.351 \\ 
3 & 1.47 & 0.141 $\pm$ 0.058 & 4.32 $\pm$ 0.946 \\ 
4 & 1.48 & 0.246 $\pm$ 0.106 & 1.79 $\pm$ 2.35 \\ 
5 & 1.46 & 0.00697 $\pm$ 0.217 & 8.67 $\pm$ 5.94 \\ 
6 & 0.46 & 0.995 $\pm$ 0.516 & -24.0 $\pm$ 16.6 \\ 
7 & 0.50 & 0.180 $\pm$ 1.41 & 5.98 $\pm$ 51.0 \\ 
\end{tabular}\caption{Fit results for standard $\chi^2$ approach with $x_{max}=2/\pi$.\label{Tab:Standard2}}
\end{center}
\end{table}

\section{Application to lattice data}

One possible application of the outlined method is the extraction of LECs in chiral perturbation theory (ChPT) from lattice data. In particular, we have studied the determination of the chiral limit value of the nucleon mass and the nucleon sigma term. There are several additional issues that need to be addressed. In these exploratory studies we again used pseudo-data generated at a set of pion mass values from the ChPT form of the nucleon mass. Our results suggest that for the naturalness prior of Eq.~(\ref{priorpdf}) larger values of $R$ are suppressed, and the main contribution to the integral over $R$ comes from the region $R\sim1-2$. It should be noted that the numerical values of the dimensionless low-energy coefficients to which the naturalness assumption applies depend on the value of the underlying scale $\Lambda$. This manifests itself in a certain sensitivity of the extracted LECs on the choice of $\Lambda$. In addition we also want to make use of detailed information on some of the parameters that appear in the ChPT expression of the nucleon mass, such as the pion decay constant and the axial coupling of the nucleon. This information allows for use of more sophisticated priors. We are continuing our investigation of these issues \cite{MPS}.

\section{Conclusions}

Extraction of the values of parameters relevant to low-energy dynamics from pertinent data is an important part of effective-field-theory calculations. We have presented a novel approach to this problem that is based on Bayesian probability theory. In this approach, prior information regarding the parameters of interest can be taken into account during the data analysis. This also allows for a more systematic inclusion of uncertainties related to truncations in the EFT. Application to a toy problem shows that our method results in an improved extraction of the low-energy constants of interest. We are continuing to study the application of these ideas to lattice QCD data and chiral perturbation theory.

\acknowledgments

We thank the organizers for a very interesting workshop and gratefully acknowledge discussions with J.~A.~McGovern. This work was supported by the US DOE under grant no. DE-FG02-93ER40756.


\begin{thebibliography}{99}

\bibitem{Schindler:2008fh}
  M.~R.~Schindler and D.~R.~Phillips,
  Annals Phys.\  {\bf 324} (2009) 682
  [Erratum-ibid.\  {\bf 324} (2009) 2051]
  [arXiv:0808.3643 [hep-ph]].


\bibitem{Sivia}
  D.~S.~Sivia, {\it Data analysis: a Bayesian tutorial}, Oxford University Press, 1996.

\bibitem{MPS}
J.~A.~McGovern, D.~R.~Phillips, M.~R.~Schindler, in preparation.

\end{thebibliography}
\end{document}